\documentclass[12pt]{article}
\usepackage{amsmath2000}
\allowdisplaybreaks
\begin{document}
\makeatletter
\@addtoreset{equation}{section}
\renewcommand{\theequation}{\thesection.\arabic{equation}}
\renewcommand{\thefootnote}{\fnsymbol{footnote}}
\def\be{\begin{equation}}
\def\ee{\end{equation}}
\def\bea{\begin{eqnarray}}
\def\eea{\end{eqnarray}}
\def\M{{\bar N}}
\def\T{{\rm T}}
\def\la{\langle}
\def\ra{\rangle}
\def\calO{{\cal O}}
\def\calN{{\cal N}}
\newcommand{\bra}[1]{\langle #1\vert}
\newcommand{\ket}[1]{\vert #1\rangle}
\newcommand{\ds}{\displaystyle}
\newcommand{\nn}{\nonumber}
\def\vac{{\rm vac}}
\begin{titlepage}
\title{
\hfill\parbox{4cm} {\normalsize CALT-68-2404\\
{\tt hep-th/0209011}}\\
\vspace{1cm}
{\bf A Note on Cubic Interactions in PP-Wave\\
Light Cone String Field Theory}
}
\author{
Peter {\sc Lee}, Sanefumi {\sc Moriyama} and Jongwon {\sc Park}
\thanks{{\tt peter, moriyama, jongwon@theory.caltech.edu}}\\
{\it California Institute of Technology 452-48,
Pasadena, CA91125, USA}\\[15pt]
}
\date{\normalsize September, 2002}
\maketitle
\thispagestyle{empty}
\begin{abstract}
\normalsize
We study the string modes in the pp-wave light-cone string field
theory. First, we clarify the discrepancy between the Neumann
coefficients for the supergravity vertex and the zero mode of the full
string one. We also repeat our previous manipulation of the prefactor
for the string modes and find that the prefactor reduces to the energy
difference of the cos modes minus that of the sin modes.
Finally, we discuss off-shell three-string processes.
\end{abstract}
\end{titlepage}

\section{Introduction}
String theory on pp-wave background and ${\cal N}=4$ SYM gauge theory
restricted to large $R$ charge were proposed to be dual in \cite{BMN}.
This proposal was partly motivated by the fact that pp-wave background
can be obtained by taking the Penrose limit of AdS space \cite{BFHP}.
The explicit comparison is made possible because string theory on the
pp-wave background can be solved \cite{M,MT} despite the fact that
there exists non-zero RR flux.
The anomalous dimension of certain gauge theory operators have been
computed in \cite{KPSS,GMR,largeN,SZ} and shown to agree with the
light-cone energy of the dual string states.
As for the string interaction part, the explicit proposal of the
correspondence between the string theory and the gauge theory
quantities \cite{largeN}
\be
\bra{1}\bra{2}\bra{3}{H}\rangle
=\mu(\Delta_1+\Delta_2-\Delta_3)C_{123},
\label{conjecture}
\ee
follows from unitarity check for large $\mu$.
There have been many reports in the literature verifying this relation
\cite{KKLP,H1,CKT1,V,LMP,SV2,Z}.
Despite the fact that all the tests for the on-shell three-point
Hamiltonian matrix elements of scalar excitations have been
successful, similar relation for vector excitations \cite{Gur,CKPRT}
or for four-point function \cite{BKPSS} seems to avert a naive
generalization.

We address three issues of the pp-wave light-cone string field
theory constructed in \cite{SV1}.
First of all, it has been noted in the literature \cite{H1} that the
supergravity Neumann matrices do not match the zero-mode of the string
Neumann matrices with $\mu\to\infty$.
In this note, we resolve the origin of this discrepancy.

The role of the prefactor in the pp-wave light-cone string field
theory was discussed in \cite{largeN,KKLP,LMP,SV2}.
Through the unitary check, the contribution of the prefactor was
proposed to be just an overall factor of difference in energy of
incoming and outgoing string states.
In \cite{LMP}, the prefactor was recast in a form to make this fact
manifest.
However, the analysis was restricted to the supergravity vertex.
In \cite{SV2}, the Hamiltonian matrix elements including the full
string prefactor were calculated as a whole and the proposal of
\cite{largeN} was confirmed to first order in $\lambda'$, but the
explicit evaluation was only restricted to two processes and the role
of the prefactor was not identified.
Here we would like to combine \cite{LMP} and \cite{SV2} and compare
the prefactor for the full string vertex with a difference in energies
of string states for any $\mu\alpha'p^+$ using a factorization
theorem of the Neumann coefficients shown in \cite{S,P}.

Thirdly, let us make an attempt to extend the conjecture to energy
non-preserving processes explicitly.
Energy non-preserving processes are proposed \cite{largeN} to
correspond to non-perturbative effects in the gauge theory side.
We shall show that the result of string theory cannot be reproduced
only from perturbative gauge theory.
We consider, at the tree-level, energy non-preserving process with two
incoming states with $m+1$ and $n+1$ impurities and one outgoing state
with $m+n$ impurities.
On the gauge theory side it is known that the contribution is
sub-leading in $1/J$ and vanishes in the pp-wave limit.
However, on the string theory side the corresponding correlation
function is proportional to 
${\bar N}^{(12)}({\bar N}^{(13)})^m({\bar N}^{(23)})^n$ and is
non-vanishing.
This is anothor sign that the proposal of \cite{largeN} should be
modified for more general string interactions.

In the following sections we shall address these questions. We
clarify the discrepancy between the supergravity vertex and the
string vertex in the following section. In sec.~3, we repeat the
manipulation of the prefactor in our previous paper \cite{LMP} for
the full string vertex. We also discuss the energy non-conserving
process in sec.~4. Finally we conclude.

\subsection*{Note added}
After our submission of the present paper to the hep-th archive, we
were informed by the author of \cite{P} that the formula (\ref{F_r^-})
in the original version of the present paper, which was first obtained
in \cite{SV2}, should be corrected by an extra factor of $i$ on the
RHS.
Accordingly, (\ref{N_-m-n}) and (\ref{nega}) also have to be corrected
by a minus sign.
Therefore, our original claim in sec.~3, that the prefactor reduces
to the energy difference, no longer holds.
Instead, it reduces to the energy difference of the non-negative
($\cos$) modes minus that of the negative ($\sin$) modes.
We have made corrections in sec.~3 accorgingly.

\section{Supergravity vertex}
In this section, we shall clarify the discrepancy between the
supergravity vertex and the zero modes of the string vertex.
In this paper, we mainly adopt the notation of \cite{SV1,SV2,KSV}.
Only in this section we set $\alpha_{(3)}=-1$ instead of introducing
$\beta$ or $y$ for simplicity. In the final result, $\alpha_{(3)}$ can
be restored on dimensional grounds.
The Neumann coefficients for the bosonic modes are given as
\be
\bar{N}^{(rs)}=\delta_{rs}1-2\sqrt{C_{(r)}}X^{(r)\T}
\frac{1}{\Gamma_a}X^{(s)}\sqrt{C_{(s)}},
\label{N^rs}
\ee
with the matrix $\Gamma_a$
\be
\Gamma_a=\sum_{r=1}^3X^{(r)}C_{(r)}X^{(r)\T}.
\label{Ga_a}
\ee
Here $X^{(r)}$ denote infinite matrices of Fourier expansion for
the third string in terms of the other two with their indices
running over the set of all integers.
If we set $X^{(3)}=1$, then $X^{(r)}$ ($r=1,2$) can be expressed as
($r,s=1,2$, $\epsilon^{12}=1$)
\be
X^{(r)}=\begin{pmatrix}
-(1/\alpha^{(r)})\sqrt{C}^{-1}A^{(r)}\sqrt{C}&0&0\\
0&1&0\\0&-(\epsilon^{rs}\alpha^{(s)}/\sqrt{2})\sqrt{C}B
&\sqrt{C}A^{(r)}\sqrt{C}^{-1}
\end{pmatrix},
\label{X^r}
\ee
with three rows(columns) representing the negative modes, the zero
mode and positive modes respectively, and $A^{(r)}_{mn}$, $B_m$,
$C_{mn}$ and $C_{(r)mn}$ are given
by
\bea
&&A^{(1)}_{mn}=(-1)^{m+n+1}\frac{2}{\pi}
\frac{\sqrt{mn}\alpha_{(1)}\sin m\pi\alpha_{(1)}}
{n^2-m^2\alpha_{(1)}^2},\\
&&A^{(2)}_{mn}=(-1)^{m}\frac{2}{\pi}
\frac{\sqrt{mn}\alpha_{(2)}\sin m\pi\alpha_{(1)}}
{n^2-m^2\alpha_{(2)}^2},\\
&&B_m=(-1)^{m+1}\frac{2}{\pi}\frac{\sin m\pi\alpha_{(1)}}
{m^{3/2}\alpha^{(1)}\alpha^{(2)}},\\
&&C_{mn}=\delta_{mn}m,\qquad
C_{(r)mn}=\delta_{mn}\omega_{(r)m},\quad r=1,2,3
\eea
with $\omega_{(r)m}=\sqrt{(\mu\alpha_{(r)})^2+m^2}$ for $r=1,2$ and
$\omega_{(3)m}=\sqrt{\mu^2+m^2}$.

If we take the large $\mu$ limit for the zero mode of the Neumann
coefficient matrices $\bar N_{00}^{(rs)}$, we find that
\be
{\bar N}_{00}^{(rs)}\rightarrow\left(\begin{matrix}
0&0&-\sqrt{\alpha_{(1)}}\\
0&0&-\sqrt{\alpha_{(2)}}\\
-\sqrt{\alpha_{(1)}}&-\sqrt{\alpha_{(2)}}&0
\end{matrix}\right).
\ee
Clearly, they do not agree with the supergravity vertex $M^{rs}$
\be
M^{rs}=\left(\begin{matrix}
\alpha_{(2)}&-\sqrt{\alpha_{(1)}\alpha_{(2)}}&-\sqrt{\alpha_{(1)}}\\
-\sqrt{\alpha_{(1)}\alpha_{(2)}}&\alpha_{(1)}&-\sqrt{\alpha_{(2)}}\\
-\sqrt{\alpha_{(1)}}&-\sqrt{\alpha_{(2)}}&0
\end{matrix}\right).
\ee

In the construction of the supergravity vertex, the dependence of
$\mu$ does not appear explicitly in the Neumann coefficient
matrices $M^{rs}$, so one might regard this mismatch as a puzzle.
However, the supergravity vertex is constructed implicitly under
the assumption that the zero modes decouple completely from the
higher ones.
This is not true in general because there are non-vanishing overlaps
between the zero modes and the positive excited modes in the Fourier
expansion matrices $X^{(r)}$.
It is only in the flat space limit $\mu\to 0$ that the zero modes
should decouple.

Let us demonstrate this observation more explicitly.
Evaluating the matrix $\Gamma_a$ (\ref{Ga_a}) by substituting the
expression for $X^{(r)}$ (\ref{X^r}), we find that the zero modes and
the positive modes decouple as
\be
\Gamma_a=\begin{pmatrix}
\sqrt{C}\Gamma_-\sqrt{C}&0&0\\
0&2\mu&0\\
0&0&\sqrt{C}\Gamma_+\sqrt{C}
\end{pmatrix}.
\ee
Using this expression, we find\footnote{The expression (\ref{N/M})
and its behavior in the flat space limit $\mu\to 0$ were also
discussed in \cite{P}.}
\be
\frac{\bar N_{00}^{(rs)}}{M^{rs}}
=1-\mu\alpha_{(1)}\alpha_{(2)}B^\T\Gamma_+^{-1}B\equiv R,
\label{N/M}
\ee
for $r,s=1,2$, while $\bar N_{00}^{(rs)}/M^{rs}=1$, for $r=3$ or
$s=3$.
{}From the above observation, we expect that $R\to 1$ as we take the
flat space limit $\mu\to 0$, while $R\to 0$ as $\mu\to\infty$.
Before proceeding to analytical computation illustrating this
behavior, let us make a few comments.

The aforementioned dependence of $R$ on $\mu$ can be seen from
numerical analysis.
In Table~\ref{R}, we present a numerical result for $R$ with
$\alpha_{(1)}=1/\sqrt{2}$ and for various value of $\mu$.
The reason we take $\alpha_{(1)}=1/\sqrt{2}$ is purely technical;
we can avoid treating the indefinite forms by adopting irrational
number for $\alpha_{(1)}$.

\begin{table}[htbp]
\begin{center}
\begin{tabular}[b]{|r|c|c|c|c|c|c|c|}
\hline
$L$~ & $\mu=1000$ & $\mu=100$ & $\mu=10$ & $\mu=1$ &
$\mu=0.1$ & $\mu=0.01$ & $\mu=0.001$ \\
\hline\hline
$10$ & $0.0469432\phantom{0}$ & $0.0471652$ &
$0.0646173$ & $0.368166$ & $0.889787$ & $0.988366$ & $0.998830$ \\
$20$ & $0.0234705\phantom{0}$ & $0.0239095$ &
 $0.0488080$ & $0.360486$ & $0.887990$ & $0.988167$ & $0.998810$ \\
$30$ & $0.0156368\phantom{0}$ & $0.0162841$ &
 $0.0446278$ & $0.358092$ & $0.887414$ & $0.988102$ & $0.998804$ \\
$40$ & $0.0117402\phantom{0}$ & $0.0125833$ &
 $0.0428047$ & $0.356936$ & $0.887132$ & $0.988071$ & $0.998801$ \\
$50$ & $0.00941868$ & $0.0104431$ &
 $0.0418053$ & $0.356258$ & $0.886966$ & $0.988053$ & $0.998799$ \\
\hline
\end{tabular}
\caption{The behavior of $R$ with $\alpha_{(1)}=1/\sqrt{2}$
for various $\mu$.}
\label{R}
\end{center}
\end{table}

Note that the second term of $R$ in (\ref{N/M}) makes $R$ deviate from
$1$, and this term comes from the off-diagonal part of $X^{(r)}$.
Since the off-diagonal part represents the overlap between the zero
modes and the positive ones, this fact confirms the reason why the two
Neumann matrices do not agree; supergravity modes do not decouple from
the string modes in general.

Let us proceed with the analytical computation.
The asymptotic behavior of $R$ in the limit $\mu\to\infty$ was
evaluated in \cite{KSV,S}. The result is
\be
R\sim\frac{1}{\pi\mu\alpha_{(1)}\alpha_{(2)}},
\ee
up to some numerical factor.
For completeness, let us also consider the flat space limit
$\mu\to 0$.
The behavior of $R$ in this limit is easily obtained by concerning an
early work of flat space light-cone string field theory \cite{GS}.
In the limit $\mu\to 0$, $\Gamma_+$ reduces to the flat space one
(called $\Gamma$ in \cite{GS}).
\be
\Gamma_+\to A^{(1)}A^{(1)\T}+A^{(2)}A^{(2)\T}+1.
\ee
Since $B^\T\Gamma^{-1}B$ was also calculated there, the behavior of
$R$ around $\mu\to 0$ simply reads
\be
R\sim 1+2\mu\bigl(\alpha_{(1)}\ln\alpha_{(1)}
+\alpha_{(2)}\ln\alpha_{(2)}\bigr).
\ee
In summary, we can reproduce the expected $\mu$-dependence of $R$
both numerically and analytically.

\section{Prefactor}
In this section, we repeat the manipulation of the prefactor in
\cite{LMP} for the full pp-wave light-cone string vertex
\cite{SV1,SV2}.
However, we will show that our previous result \cite{LMP} for the
supergravity modes does not hold for the full string vertex;
the original expectation that the prefactor reduces to the energy
difference between the incoming and outgoing string states is no
longer true.
Instead, it reduces to the energy difference of the non-negative
($\cos$) modes minus that of the negative ($\sin$) modes.

The prefactor in the oscillator basis is given \cite{SV2} as
$K^I\tilde K^J v_{IJ}(\Lambda)$, where $K=K_++K_-$ and
$\tilde K=K_+-K_-$ with
\be
K_+=\sum_{r=1}^3\sum_{m=0}^\infty F_{(r)m}^+a_{m(r)}^\dagger,
\qquad
K_-=\sum_{r=1}^3\sum_{m=1}^\infty F_{(r)m}^-a_{-m(r)}^\dagger,
\ee
and ($\alpha=\alpha_{(1)}\alpha_{(2)}\alpha_{(3)}$)
\bea
&&v^{IJ}=\delta^{IJ}
-\frac{i}{\alpha}\gamma^{IJ}_{ab}\Lambda^a\Lambda^b
+\frac{1}{6\alpha^2}\gamma^{IK}_{ab}\gamma^{JK}_{cd}
\Lambda^a\Lambda^b\Lambda^c\Lambda^d\nn\\
&&\qquad
-\frac{4i}{6!\alpha^3}\gamma^{IJ}_{ab}\epsilon_{abcdefgh}
\Lambda^c\Lambda^d\Lambda^e\Lambda^f\Lambda^g\Lambda^h
+\frac{16}{8!\alpha^4}\delta^{IJ}\epsilon_{abcdefgh}
\Lambda^a\Lambda^b\Lambda^c\Lambda^d
\Lambda^e\Lambda^f\Lambda^g\Lambda^h.
\label{v^IJ}
\eea
When one restricts to the bosonic excitation, only the third term in
(\ref{v^IJ}) contributes.
In addition, from the structure of the gamma matrices \cite{LMP,KKLP},
we know that except for a relative minus sign\footnote{See also
\cite{CKPRT} for a recent proposal related to this relative minus
sign.} between the two SO(4)'s, the gamma matrix reduces to a
Kronecker delta.
Hence, the prefactor is given explicitly as
\be
\biggl(\sum_{r=1}^3\sum_{m=0}^\infty
F_{(r)m}^+a_{m(r)}^\dagger\biggr)^2
-\biggl(\sum_{r=1}^3\sum_{m=1}^\infty
F_{(r)m}^-a_{-m(r)}^\dagger\biggr)^2.
\label{prefac}
\ee
On the other hand, the difference in energy is given by
\be
P^-_1+P^-_2-P^-_3=\sum_{r=1}^3\sum_{m=-\infty}^\infty
\bigl(\omega_{(r)m}/\alpha_{(r)}\bigr)a_{m(r)}^\dagger a_{m(r)}.
\ee
Acting it on the bosonic vertex $E_a\ket{\vac}$ gives
\bea
\bigl(P^-_1+P^-_2-P^-_3\bigr)E_a\ket{\vac}
=\sum_{r,s=1}^3\sum_{m,n=0}^\infty
a_{m(r)}^\dagger\bigl(\omega_{(r)m}/\alpha_{(r)}\bigr)
{\bar N}^{(rs)}_{mn}a_{n(s)}^\dagger
E_a\ket{\vac}\phantom{.}&&\nn\\
+\sum_{r,s=1}^3\sum_{m,n=1}^\infty
a_{-m(r)}^\dagger\bigl(\omega_{(r)m}/\alpha_{(r)}\bigr)
{\bar N}^{(rs)}_{-m-n}a_{-n(s)}^\dagger
E_a\ket{\vac}.&&
\label{preserve}
\eea

Let us compare this expression (\ref{preserve}) with the prefactor
(\ref{prefac}).
We first concentrate on the positive modes.
For these modes, $F_{(r)m}^+$ is given as \cite{SV2}
\be
F_{(r)}^+=\frac{1}{\sqrt{2}}\frac{\alpha}{\alpha_{(r)}}
\sqrt{CC_{(r)}}U_{(r)}^{-1}A^{(r)\T}\Upsilon^{-1}B,
\label{F_r^+}
\ee
up to normalization\footnote{In \cite{SV2} the overall normalization
of the prefactor was fixed by comparing with the supergravity vertex
$M^{rs}$ with $r,s=1,2$. This normalization is reliable only when
$\mu$ is small.} with
\be
\Upsilon=\sum_{r=1}^3A^{(r)}U_{(r)}^{-1}A^{(r)\T},
\qquad U_{(r)}=(C_{(r)}-\mu\alpha_{(r)})/C,
\ee
and the Neumann coefficients are shown to have the following useful
factorization property \cite{S,P}:
\be
{\bar N}^{(rs)}_{mn}=-\frac{\alpha}{R}
\frac{m{\bar N}^{(r)}_m{\bar N}^{(s)}_nn}
{\alpha_{(s)}\omega_{m(r)}+\alpha_{(r)}\omega_{n(s)}},
\label{decomp}
\ee
with
\be
{\bar N}^{(r)}=-\sqrt{\frac{C_{(r)}}{C}}
U_{(r)}^{-1}A^{(r)\T}\Gamma_+^{-1}B.
\ee
Since we can also show the property $\Upsilon^{-1}B=\Gamma_+^{-1}B/R$
\cite{S,P}, ${\bar N}^{(r)}_m$ is closely related to $F_{(r)m}$.
In fact we can rewrite the factorization theorem (\ref{decomp}) in
terms of $F_{(r)}$ as
\be
\bar N^{(rs)}_{mn}=-\frac{R}{\alpha}
\frac{2}{\omega_{m(r)}/\alpha_{(r)}+\omega_{n(s)}/\alpha_{(s)}}
F_{m(r)}^+F_{n(s)}^+.
\label{decompose}
\ee
Although the expression for $F_{(r)}^+$ (\ref{F_r^+}) and the
factorization theorem (\ref{decomp}) were originally obtained for the
positive modes, one can show that the present formula
(\ref{decompose}) holds also for the zero mode if the indefinite from
of ${\bar N}^{(13)}_{00}$ and ${\bar N}^{(23)}_{00}$ is interpreted
properly.
Substituting this factorization theorem (\ref{decompose}) into the
energy difference (\ref{preserve}) and exchanging the dummy labels
$(r,m)$ and $(s,n)$, we find
\be
\sum_{r,s=1}^3\sum_{m,n=0}^\infty
a_{m(r)}^\dagger\bigl(\omega_{m(r)}/\alpha_{(r)}\bigr)
{\bar N}^{(rs)}_{mn}a_{n(s)}^\dagger
E_a\ket{\vac}
=-\frac{R}{\alpha}
\biggl(\sum_{r=1}^3\sum_{m=0}^\infty
F_{m(r)}^+a_{m(r)}^\dagger\biggr)^2E_a\ket{\vac}.
\label{posi}
\ee
We can also repeat the above calculation for the negative modes by
noting\footnote{We are grateful to A. Pankiewicz for informing us that
(\ref{F_r^-}) in the original version, which was first obtained in
\cite{SV2}, should be corrected by an extra factor $i$.} \cite{SV2}
\bea
&&{\bar N}^{(rs)}_{-m-n}
=-\bigl(U_{(r)}{\bar N}^{(rs)}U_{(s)}\bigr)_{mn},\\
&&F_{(r)}^-=iU_{(r)}F_{(r)}^+\label{F_r^-}.
\eea
Using these formulae, the Neumann coefficient matrices for the
negative modes can be expressed as
\be
{\bar N}^{(rs)}_{-m-n}=-\frac{R}{\alpha}
\frac{2}{\omega_{m(r)}/\alpha_{(r)}+\omega_{n(s)}/\alpha_{(s)}}
F_{m(r)}^-F_{n(s)}^-.\label{N_-m-n}
\ee
Therefore, the contribution from the negative modes gives
\be
\sum_{r,s=1}^3\sum_{m,n=1}^\infty
a_{-m(r)}^\dagger\bigl(\omega_{m(r)}/\alpha_{(r)}\bigr)
{\bar N}^{(rs)}_{-m-n}a_{-n(s)}^\dagger
E_a\ket{\vac}=-\frac{R}{\alpha}
\biggl(\sum_{r=1}^3\sum_{m=1}^\infty
F_{m(r)}^-a_{-m(r)}^\dagger\biggr)^2E_a\ket{\vac}.
\label{nega}
\ee

Consequently, the prefactor does not reduce to the energy difference,
but reduces to the energy difference of the non-negative ($\cos$)
modes minus that of the negative ($\sin$) modes:
\be
({\rm Prefactor})\sim
(P_1^-+P_2^--P_3^-)\bigr\vert_{\cos}
-(P_1^-+P_2^--P_3^-)\bigr\vert_{\sin}.
\ee
Note that the prefactor is diagonal only in the cos/sin basis, and not
in the exp basis which is natural in the context of the PP-wave/SYM
correspondence.
Here $\sim$ means that this relation holds up to some scalar factor,
because the normalization of the prefactor is still unknown.
As we pointed out in the footnote of (\ref{F_r^+}), in \cite{SV2} the
normalization of the prefactor was determined by comparing with the
supergravity vertex $M^{rs}$ ($r,s=1,2$) and this normalization is
reliable only for small $\mu$.
Therefore, the overall normalization should be fixed in another way.
For example, if we simply replace $M^{rs}$ by the zero modes of the
string Neumann matrices ${\bar N}^{(rs)}_{00}=M^{rs}R$ when fixing the
normalization, then the scalar factor no longer depends on $\mu$
but only on some numbers and $\alpha$.
To be more precise, it is necessary to determine the overall scalar
factor completely without mentioning to the supergravity vertex.
This issue of overall normalization constant can be circumvented by
computing ratio of three-point functions as done in \cite{H1,CKT1}.

\section{Towards energy non-preserving process}
Having acquired a systematic viewpoint of the prefactor, let us
proceed by checking if the string/gauge correspondence holds
beyond the energy conserving processes.
We shall consider the process with two incoming states with $m+1$ and
$n+1$ impurities and one outgoing state with $m+n$ impurities.
Here we shall restrict ourselves to the zero modes and abbreviate
${\bar N}^{(rs)}_{00}$ as ${\bar N}_{rs}$.
First of all, let us consider the string theory side.
Using \cite{LMP} and the argument of sec.~3, all we have to do is to
calculate the following quantity:
\be
\la a_1^{m+1}a_2^{n+1}a_3^{m+n}\ra,
\ee
with $\la{\cal O}\ra$ defined as
\be
\la{\cal O}\ra\equiv\bra{\vac}{\cal O}E_a\ket{\vac}.
\ee
Since $\M_{33}=0$, $a_3$ cannot be contracted with itself.
Therefore we have three types of terms:
$\M_{11}\M_{13}^{m-1}\M_{23}^{n+1}$,
$\M_{22}\M_{13}^{m+1}\M_{23}^{n-1}$ and $\M_{12}\M_{13}^m\M_{23}^n$.
The combinatorial coefficient of $\M_{11}\M_{13}^{m-1}\M_{23}^{n+1}$ is
obtained as follows. First of all, since $a_2$ is always contracted
with $a_3$, we have $(m+n)(m+n-1)\cdots m$ ways to do this.
The rest of $a_3$ have to be contracted with $a_1$, and there are
$(m+1)m\cdots 3$ ways to do this. Finally, the remaining two $a_1$'s
have to be contracted by themselves uniquely.
Therefore, the coefficient of $\M_{11}\M_{13}^{m-1}\M_{23}^{n+1}$ is
given as
\be
(m+n)(m+n-1)\cdots m\cdot(m+1)m\cdots 3\cdot 1
=\frac{(m+1)m}{2}(m+n)!.
\ee
Similar reasoning yields the coefficient of
$\M_{22}\M_{13}^{m+1}\M_{23}^{n-1}$ to be
\be
(m+n)(m+n-1)\cdots n\cdot(n+1)n\cdots 3\cdot
1=\frac{(n+1)n}{2}(m+n)!.
\ee
The coefficient of $\M_{12}\M_{13}^m\M_{23}^n$ can be computed by
substracting the previous two cases from the combinatoric factor of
contracting all $a_3$ with $a_1$ or $a_2$.
The coefficient of $\M_{12}\M_{13}^m\M_{23}^n$ is found to be
\bea
&(m+n+2)(m+n+1)\cdots 3-{\ds\frac{(m+1)m}{2}}(m+n)!
-{\ds\frac{(n+1)n}{2}}(m+n)!&\nn\\
&\qquad=(m+1)(n+1)(m+n)!.&
\eea
To summarize, the correctly normalized matrix element is given as
\bea
&\ds\bigg\la\frac{a_1^{m+1}}{\sqrt{(m+1)!}}
\frac{a_2^{n+1}}{\sqrt{(n+1)!}}
\frac{a_3^{m+n}}{\sqrt{(m+n)!}}\bigg\ra
=\frac{(m+n)!}{\sqrt{(m+1)!(n+1)!(m+n)!}}\times&\nn\\
&\ds\biggl\{\frac{(m+1)m}{2}\M_{11}\M_{13}^{m-1}\M_{23}^{n+1}
+\frac{(n+1)n}{2}\M_{22}\M_{13}^{m+1}\M_{23}^{n-1}
+(m+1)(n+1)\M_{12}\M_{13}^m\M_{23}^n\biggr\}.&
\label{str}
\eea

Now let us turn to the gauge theory side.
At the tree-level, we have
\be
\la\calO_{0**(m+1)}^{J_1}\calO_{0**(n+1)}^{J_2}
\bar\calO_{0**(m+n)}^J\ra
=\frac{1}{\calN_{J,m+n}}
\frac{1}{\calN_{J_1,m+1}}
\frac{1}{\calN_{J_2,n+1}}
\frac{(J_1+m)!}{J_1!m!}
\frac{(J_2+n)!}{J_2!n!},
\ee
with $\calN_{J,n}=\sqrt{N^{J+n}(J+n-1)!/(J!n!)}$.
To compare this result with that of the string theory side, we have to
take the pp-wave limit: $J,N\to\infty$ with $J^2/N$ fixed.
In this limit, the ratio to the vacuum three-point function is given
by
\bea
\frac{\la\calO_{0**(m+1)}^{J_1}\calO_{0**(n+1)}^{J_2}
\bar\calO_{0**(m+n)}^J\ra}
{\la\calO^{J_1}\calO^{J_2}\bar\calO^J\ra}
\to\sqrt{(m+1)(n+1)}\sqrt{\frac{(m+n)!}{m!n!}}\frac{1}{J}
\biggl(\frac{J_1}{J}\biggr)^{(m-1)/2}
\biggl(\frac{J_2}{J}\biggr)^{(n-1)/2}.
\label{gau}
\eea
Therefore, in the pp-wave limit the perturbative field theory results
simply vanish at the tree-level.
Next order in perturbation theory would give a contribution of order
$\lambda$, but due to the usual non-renormalization theorem for two
and three point functions of chiral primary operators \cite{LMRS,DFS},
we do not expect any perturbative corrections to above amplitudes.

{}From the analysis done in sec.~2, we know that $\M_{rs}$ scales as
$1/\mu$ for large $\mu$ for $r,s=1,2$.
Hence, the string amplitude scales as half-integer power of the
effective coupling $\lambda'$ at small $\lambda'$.
It seems difficult to reproduce this behavior in perturbative gauge
theory, and in order to reproduce the string theory results, we need
to include non-perturbative effects as well \cite{KSV}.
The similarity between the coefficient of (\ref{gau}) and that of the
last term of (\ref{str}) might be a clue for resolving this mismatch.

\section{Conclusion}
In this note, we have reexamined the pp-wave light-cone string field
theory.
In doing so, we resolved the apparent puzzle regarding the mismatch
between supergravity Neumann matrices and the fully string ones.
The mismatch is shown to be due to the overlap of the zero modes with
the excited ones.
The match is of course restored in the flat space limit $\mu\to 0$.
Following this\footnote{This conclusion has been changed from the
original version.}, we concluded that the full string prefactor does
not reduce to the difference in energy between the incoming and
outgoing string states.
Instead, it reduces to the energy difference of the non-negative
($\cos$) modes minus that of the negative ($\sin$) modes.
Finally, we showed that the proposal of \cite{largeN} does not naively
generalize to the energy non-preserving amplitudes.
We expect that non-perturbative effects play an important role here.

\subsection*{Acknowledgment}
We would like to thank Jaume Gomis and Takuya Okuda for useful
discussions and comments.
This research was supported in part by DOE grant DE-FG03-92-ER40701.
S.~M.~was supported in part by JSPS Postdoctoral Fellowships for
Research Abroad H14-472.

\end{document}